\documentclass[10pt,conference]{IEEEtran}

\usepackage{cite}
\usepackage{amsmath,amssymb,amsfonts}
\usepackage{graphicx}
\usepackage{textcomp}
\usepackage{xcolor}
\usepackage{algorithm}
\usepackage{hyperref}
\usepackage{orcidlink}
\usepackage{subcaption}
\usepackage{algpseudocode}
\usepackage{mathtools}
\usepackage{comment}

\usepackage{array} 
\usepackage{tabularx}
\def\BibTeX{{\rm B\kern-.05em{\sc i\kern-.025em b}\kern-.08em
    T\kern-.1667em\lower.7ex\hbox{E}\kern-.125emX}}

\newcommand\change[1]{\textcolor{red}{#1}}

\DeclareMathOperator*{\MCX}{MCX}
\DeclareMathOperator*{\CR}{CR}

\begin{document}

\title{Autocallable Options Pricing with Integration-Based Exponential Amplitude Loading}
\author{
    \IEEEauthorblockN{
    Francesca Cibrario\orcidlink{0009-0007-8290-4992}\IEEEauthorrefmark{1},
    Ron Cohen\IEEEauthorrefmark{2}, 
    Emanuele Dri\orcidlink{0000-0002-5144-1514}\IEEEauthorrefmark{3}, 
    Christian Mattia\orcidlink{0009-0001-6546-6341}\IEEEauthorrefmark{1}, 
    Or Samimi Golan\IEEEauthorrefmark{2},
    Tamuz Danzig\IEEEauthorrefmark{2}, \\
    Giacomo Ranieri\orcidlink{0009-0005-0488-2121}\IEEEauthorrefmark{1}, 
    Hanan Rosemarin\IEEEauthorrefmark{2}, 
    Davide Corbelletto\orcidlink{0009-0003-8830-2619}\IEEEauthorrefmark{1}, 
    Amir Naveh\IEEEauthorrefmark{2},
    Bartolomeo Montrucchio\IEEEauthorrefmark{4}
    }
    \\
    \IEEEauthorblockA{
    \IEEEauthorrefmark{1}Data, AI and Technology, Intesa Sanpaolo, Torino, Italy\\
    francesca.cibrario@intesasanpaolo.com, 
    christian.mattia@intesasanpaolo.com,\\
    giacomo.ranieri@intesasanpaolo.com,
     davide.corbelletto@intesasanpaolo.com\\
    \IEEEauthorrefmark{2}Classiq Technologies, Tel Aviv, Israel\\
    ron@classiq.io, orsa@classiq.io, tamuz@classiq.io, hanan@classiq.io, amir@classiq.io\\
    \IEEEauthorrefmark{3}Fondazione LINKS, Torino, Italy\\
    emanuele.dri@linksfoundation.com\\
    \IEEEauthorrefmark{4}DAUIN, Politecnico di Torino, Torino, Italy\\ bartolomeo.montrucchio@polito.it
    }
}

\maketitle

\thispagestyle{plain}
\pagestyle{plain}

\begin{abstract}
We present a comprehensive quantum algorithm tailored for pricing autocallable options, offering a full implementation and experimental validation.
Our experiments include simulations conducted on high performance computing (HPC) hardware, along with an empirical analysis of convergence to the classically estimated value.
Our key innovation is an improved integration‐based exponential amplitude loading technique that reduces circuit depth compared to state-of-the-art approaches. 
A detailed complexity analysis in a relevant setting shows a $\sim$50x reduction in T-depth for the payoff component relative to previous methods. 
These contributions represent a step toward more efficient quantum approaches to pricing complex financial derivatives.

\end{abstract}


\section{Introduction}
Quantum computing is widely expected to benefit several specific types of computationally intensive problems.
Derivative pricing, along with other tasks in the financial sector, is an excellent candidate to test the potential and limitations of the quantum computing paradigm \cite{Orus2019,Egger2020,Herman2023,Sotelo2024}.
Various studies and contributions focus on the implementation and analysis of quantum algorithms to price different types of structured products.

Several existing proposals hint at the possibility of achieving a quantum advantage over the traditional Monte Carlo approach in the asymptotic regime for \textit{path-independent} financial products, such as European options. 
Although these algorithms constitute a fundamental first step toward the study and analysis of more complex derivatives, these products do not necessarily present a significant challenge for traditional pricing methodologies.
There is room to explore genuine practical utility in the design of quantum algorithms to price more complex classes of products, such as \textit{path-dependent} derivatives. 
Autocallable options, which are path-dependent financial products, are ideal for quantum experiments. They have already been used for benchmarking in \cite{chakrabarti2021, Stamatopoulos2024derivativepricing} to assess the potential and requirements to achieve quantum advantage in this area. 

Specifically, Chakrabarti et al. \cite{chakrabarti2021} proposed a novel re-parameterization method that integrates pre-trained variational circuits with fault-tolerant quantum computing to lower the amount of resources needed for achieving quantum advantage. 
More recently, work by Stamatopoulus and Zeng \cite{Stamatopoulos2024derivativepricing} focused on limiting the quantum arithmetic components needed for derivative pricing on quantum computers. 
To avoid quantum arithmetic, their methodology leverages Quantum Signal Processing (QSP) by encoding the derivative's payoff directly into quantum amplitudes. This approach, effectively diminishes the requirements for achieving quantum advantage.
Cibrario et al. \cite{cibrario2024} proposed an algorithm to price rainbow options that reduces the use of quantum arithmetic through an alternative approach to QSP. This decreases the depth of the module that computes the payoff as an amplitude but affects the normalization factor required for the entire algorithm.
We propose an enhancement to their exponential amplitude loading methodology to build a novel end-to-end algorithm for pricing autocallable options. The algorithm offers improvements in the depth of the corresponding quantum circuit relative to the state-of-the-art. 
In addition, we built quantum circuits to test our algorithm using the Qmod language \cite{Qmod_docs}.  
We simulated the circuits for a small instance of the autocallable pricing problem to validate the results against a classical benchmark. We conducted experiments with up to 33 qubits using the LEONARDO pre-exascale supercomputer\cite{Turisini2024}.
The proposed methodology and results represent a step towards the efficient pricing of complex derivative products, which is essential to achieve \textit{quantum utility}, a practical advantage of quantum computing over classical methods. This challenge remains one of the most significant in the field today.

The remainder of this paper is structured as follows. Section \ref{sec:quantum_derivative_pricing} provides a basic overview of derivative pricing (with a focus on autocallable options) and the related quantum methods existing in the literature. 
Section \ref{sec:methodology} describes the novel implementation that constitutes the main contribution of this work, while Section \ref{sec:complexity_analysis} analyzes the complexity of the building blocks of our method. 
Then, Section \ref{sec:experiments} presents the results of an extensive simulation campaign aimed at validating the proposed approach and illustrating the effects of scaling. 
Finally, Section \ref{sec:conclusions} summarizes the main outcomes of this work and illustrates potential future avenues for further development.

\section{Quantum Derivative Pricing}
\label{sec:quantum_derivative_pricing}
A derivative is a financial contract in which two parties agree on a future transaction. This kind of contract depends on the value of its underlying variables, such as stocks, commodities, currencies, or market indexes.
Some derivatives, such as forwards and futures, impose the obligation to buy or sell the underlying variables. Options grant the right to buy or sell, but not the obligation \cite{hull2009}. 
Determining the price of options is a crucial task for financial institutions. 

An option can be a \textit{call} or a \textit{put}, depending on whether it grants the right to buy or sell the underlying variables by a certain date (i.e., \textit{maturity date}) and at a given price (i.e., \textit{strike price}). 
European options allow the contract owner to exercise this right only upon the maturity date, while American options can also be bought or sold before that specific point in time. 
An option can be defined based on a single underlying variable (\textit{single-asset}) or on multiple variables (\textit{multi-asset}). 
The payoff is the profit obtained by the option holder when the contract ends. If exercising the option is not advantageous, the payoff is zero and the option is considered \textit{out of the money}; otherwise, the payoff is positive and the option is said to be \textit{in the money}.
The payoff can depend solely on the value of the asset's price on the maturity date (\textit{path-independent option}) or on the path the asset price follows over the entire life of the contract (\textit{path-dependent option}).

Pricing an option means assigning a fair value to the contract by estimating the expected discounted payoff. Discounting a future payoff means evaluating its present value. Given a payoff $f$ at time $t$, its value today is given by $f \cdot e^{-rt}$, where $r$ is the risk-free interest rate.

Multi-asset path-dependent options are challenging to price, since the calculation must be performed by modeling the distribution of correlated asset prices over multiple time steps. 

Different models can be adopted to represent these distributions. One such example is the Black-Scholes model, which assumes the following dynamics for the evolution of asset price:
\begin{equation}
\label{eqn:GBM}
dS_t = \hat{\mu}S_tdt + \hat{\sigma} S_tdW_t
\end{equation}
where $S_t$ is the asset price at time $t$, $dW_t$ is a Wiener process,  $\hat{\mu}$ is the drift, and $\hat{\sigma}$ is the volatility \cite{hull2009}. In this type of model, $\hat{\sigma}$ is constant, while in other more realistic models, it is considered stochastic or local.
The parameters of the Black-Scholes model can be estimated easily by observing historical data, while local and stochastic volatility models require calibration. Only under certain circumstances is it possible to price an option in a closed form by solving the stochastic differential equation or determining an analytical solution.

Monte Carlo simulation can be used as an alternative method to estimate the expected discounted payoff, especially when a closed form does not exist.  It constructs different paths (scenarios) for asset price evolution over discretized time steps. Specifically, the asset price evolution of Equation \ref{eqn:GBM} can be discretized as:
\begin{equation}
\label{eqn:S(t)}
S(t+ \Delta t) = S(t) e^{\mu\Delta t + {\sigma} \alpha \sqrt{\Delta t}}
\end{equation}

where $\mu = \hat{\mu}-\frac{\hat{\sigma}^2}{2}$, $\sigma=\hat{\sigma}$, and $\alpha$ is a random extraction from a standard normal distribution (asset prices are log-normally distributed). Different scenarios can be built by sampling different random paths. Once the value of the asset price is known for all time steps, the discounted payoff can be calculated for each scenario and then averaged across all paths to estimate the expected payoff.  
The convergence rate of this classical method is $1/\sqrt{M}$, where $M$ is the number of scenarios simulated.

Quantum computing promises a quadratic speed-up compared to the Monte Carlo approach by leveraging the Quantum Amplitude Estimation (QAE) algorithm \cite{Montanaro2015}. 
The quantum approach provides a convergence rate of $1/N$, where $N$ is the number of calls to the Grover operator, which is computationally analogous to a Monte Carlo path evaluation.
Building on this theoretical result, several studies have focused on the pricing problem, with the Black-Scholes model serving as a primary framework. Rebentrost, Gupt and Bromley \cite{Rebentrost2018} presented a quantum approach for pricing single-asset derivatives. 
Other subsequent works have addressed the challenge of extending the range of options that can be priced using quantum computing, while reducing the complexity of each sub-module in the quantum algorithm to fully exploit the quadratic speed-up.
Stamatopoulos et al. \cite{Stamatopoulos2020} analyzed a wide range of derivatives, including multi-asset and path-dependent ones, loading the log-normal distribution of asset prices into a quantum state. 
Chakrabarti et al. \cite{chakrabarti2021} provided an estimate of the resources required to achieve quantum advantage in derivative pricing. They also proposed an optimization of the probability distribution encoding module by loading the log-returns distribution instead of the asset prices one. This approach introduced the need to transition from the log-return space to the price space using quantum arithmetic.
In \cite{Stamatopoulos2024derivativepricing} and \cite{cibrario2024}, two strategies were proposed to delay the transition and avoid the need for quantum arithmetic. Other studies have focused on extending the quantum pricing approach beyond the Black-Scholes model, considering models with local volatility \cite{Kaneko_Miyamoto_Takeda_Yoshino_2022} or stochastic volatility \cite{Wang2024optionpricingunder}, without addressing the calibration phase.

\begin{figure}[ht]
    \centering
    \begin{subfigure}{1\linewidth}
        \centering
        \includegraphics[trim={40 0 0 0}, width=\linewidth]{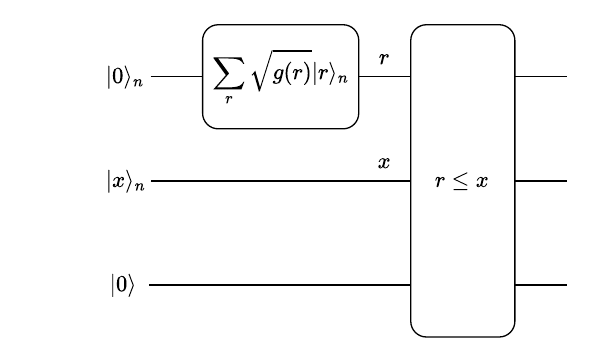}
        \caption{Circuit implementing the integration amplitude loading. The whole circuit is composed of two sub-modules: (1) a state preparation loading a $g$ function into the amplitude and (2) a comparator. The comparator ($r\leq x$) behaves like an integrator of the $g$ function with $x$ as its limit. The result of the integral is collected in the amplitude of the $|1\rangle$ state of the target qubit (i.e., the qubit storing the outcome of the comparison).}
        \label{fig:integrator}
    \end{subfigure}
    \hfill
    \begin{subfigure}{1\linewidth}
        \centering
        \includegraphics[trim={49 0 0 0}, width=\linewidth]{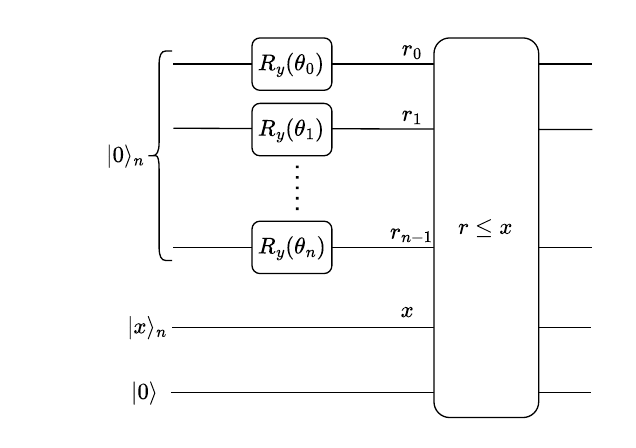}
        \caption{Circuit implementing the integration amplitude loading for the exponential function reported in Equation \ref{eqn:exponential_integrator}. It consists in implementing the exponential state preparation  with $n$ parallel $R_y$ with rotation angles defined in Equation \ref{eq:sp_exp_angles}.}
        \label{fig:exponential_sp}
    \end{subfigure}
    \caption{
    Integration amplitude loading circuit for a generic function (\protect\subref{fig:integrator}) and for the exponential one (\protect\subref{fig:exponential_sp}).}
    \label{fig:subfigures}
\end{figure}

\subsection{Autocallable Options}
\label{subsec:autocallable_sa}
An \textit{automatically callable} (or autocallable) option is a structured financial derivative instrument designed to provide potential payouts based on specific conditions met during its lifespan.
Autocallable options are popular in structured finance and computationally expensive to price: this provides strong motivation to explore the potential of quantum computing in pricing such options.
Autocallable options are characterized by the following features:

\begin{itemize}
    \item \textbf{Single or Multi-asset:} Their performance can be tied to a single underlying asset or a basket of assets. With multiple assets, one of them can be used as the reference for payoff calculation; generally, this will be either the best-performing (\textit{best-of}) or worst-performing (\textit{worst-of}) asset.
    \item \textbf{Path-dependency:} The asset has predefined dates when its performance is evaluated, creating multi-step observation periods (e.g., monthly or quarterly) that represent opportunities to trigger a payout.
    \item \textbf{Notional Value:} Autocallables are typically defined in terms of asset returns rather than prices, and, therefore, in terms of strike returns instead of strike prices. Payouts are calculated as a fraction of a notional amount, $V$, which determines the contract size in monetary units (it may be equal to the asset's initial price).
    \item \textbf{Binary Payoffs:} The derivative includes a sequence of binary options. A binary option's payoff is a fixed amount if a condition is met; otherwise, the payoff is $0$. These options, in the context of autocallable derivatives, are linked to specific dates within the multi-step observation period. Each $i$-th binary option has an expiration date $t_i$ upon which the condition is checked and the payoff is computed. If the asset return $r_{t_i}$ at $t_i$, exceeds a strike return $k_i$, the binary option is considered in the money and pays out a fixed amount $f_i$. This payout can also be expressed as a percentage of the notional value. When a binary option triggers a payout, the contract is immediately terminated. 
    \item \textbf{Short Knock-in Put:} Along with the binary options, the derivative contract includes a short knock-in put option that expires on maturity date $T$. Since the autocallable contract terminates if any binary option is in the money, the put option is only evaluated if none of the binary options have provided a payout. The payoff of the put option depends on the difference between the asset return and the strike return $K$. Due to its knock-in nature, one condition for a non-zero payoff is that the asset return $r_t$ crosses a barrier $b$ at least once during the option’s lifetime. The short position means that the option holder must pay the issuer if the payoff is non-zero. The payoff related to the put option is given by:
    $$
        f=
        \begin{cases}
        V\left(r_T-K\right), & \text {if } r_T<K \text { and } r_t<b,   \forall t \in[0,T] \\
        0, & \text {otherwise}
        \end{cases}
    $$
    where $r_T$ the return at the final time $T$.
\end{itemize}

\subsection{Quantum Approach}
\label{subsec:quantumoptionpricing_sa}
The quantum approaches for derivative pricing usually focus on two steps: probability distribution encoding and payoff evaluation. 
First, a probability distribution over discretized asset paths is loaded into a quantum register through a state preparation operator. Let \(\omega\) denote a discretized path. This paths' state preparation operator creates the following state:

\begin{equation}
\sum_{\omega} \sqrt{p(\omega)}\, |\omega\rangle,
\end{equation}

where \(p(\omega)\) is the probability of path \(\omega\).
Second, the discounted payoff function rescaled between $0$ and $1$ ($\tilde{f}$) is calculated and encoded in the amplitude of a target qubit. 

\begin{equation}
\label{eqn:payoff_ae}
\begin{split}  
&\sum_\omega \sqrt{\tilde{f}\left(\omega\right)p\left(\omega\right)}\left|\omega\right\rangle|1\rangle +\\
&\sum_\omega \sqrt{(1-\tilde{f}\left(\omega\right)) p\left(\omega\right)}\left|\omega\right\rangle|0\rangle
\end{split}
\end{equation}

The probability of measuring the target qubit in the state \(|1\rangle\) is:

\begin{equation}
E[\tilde{f}] = \sum_{\omega} p(\omega) \tilde{f}(\omega)
\end{equation}
which is the rescaled expected discounted payoff of the option. 

Through Quantum Amplitude Estimation (QAE), an estimation of $\mathbb{E}[\tilde{f}]$ can be performed with a query complexity of \(O(1/\epsilon)\), where \(\epsilon\) is the desired error tolerance. This step ultimately provides a quadratic speedup compared to the classical Monte Carlo complexity of $O(1/\epsilon^2)$. The QAE technique requires the use of an additional quantum register to encode the estimate. However, certain extensions of the methodology have been developed to circumvent this requirement. A notable example is the Iterative Quantum Amplitude Estimation (IQAE) as discussed in \cite{Grinko2021}.

Regarding the paths' state preparation, the \textit{re-parameterization method} proposed in \cite{chakrabarti2021} exploits the fact that log-returns follow a normal distribution. 
They used variationally pre-trained Gaussian loaders ($\mathit{G}$) to load the standard normal distribution:

\begin{equation}
G: |0\rangle \longrightarrow \sum_{g=0}^{2^n-1} \sqrt{p(g)}\, |g\rangle
\end{equation}

where \(p(g)\) are the discretized probabilities of a standard normal distribution (with $2^n$ discretization points).
Affine transformations are then used to obtain the desired mean and standard deviation, exploiting the following reformulation of Equation \ref{eqn:S(t)}: 
\begin{equation}
\label{eqn:ln(S(t))}
\hat{l}(t) = \ln{\frac{S(t+ \Delta t)}{S(t)}} = \mu \Delta t + \sigma \alpha \sqrt{\Delta t}
\end{equation}
Using this approach, they had to apply quantum arithmetic to compute log-returns by accumulation $l(t) = \sum_{x=0}^t{\hat{l}(x)}$ and then transition from the log-return space to return space $r(t) = e^{l(t)}$.

To avoid the use of expensive quantum arithmetic, techniques such as Quantum Signal Processing (QSP) have been applied to create a unified payoff component that handles both the transition from log-return space to return space and the direct encoding of the payoff function into quantum amplitudes\cite{Stamatopoulos2024derivativepricing}.
To compute a non-constant payoff, the state is first prepared using a unitary operator $U_{sqrt}$ that encodes a scalar number \(u\) into an amplitude:

\begin{equation}
U_{sqrt}\, |u\rangle |0\rangle \longrightarrow |u\rangle \left(\sqrt{u}\, |\psi_0\rangle |0\rangle + \sqrt{1-u}\, |\psi_1\rangle |1\rangle\right).
\end{equation}

QSP is then used to apply a polynomial transformation to encode an approximation of the function in the amplitude of the QAE target state:

\begin{equation}
\tilde{f}(u) = \sqrt{\frac{A e^{u} - B}{C}},
\end{equation}

where the constants \(A\), \(B\), and \(C\) are used to compute the payoff and to ensure \(\tilde{f}(u) \in [0,1]\).

In the context of autocallable options, the authors claimed to reduce the logical clock rate needed for quantum advantage by a factor of $\sim 5$x with respect to \cite{chakrabarti2021}. 
Moreover, they claim a reduction of the required T-gates by $\sim 16$x and of the necessary logical qubits by $\sim 4$x. 



In this work, we propose an alternative method that requires fewer resources than the QSP-based approach. We enhance the integration-based exponential amplitude loading methodology introduced in \cite{cibrario2024} (detailed in Section \ref{subsec:intal}) to avoid degrading the rescaling factor needed in the post-processing operations.

\section{Methodology}
\label{sec:methodology}
We propose a quantum algorithm for pricing autocallable options (path-dependent derivatives) and provide a concrete implementation for the pricing of a single asset while generalizing the complexity analysis for the multi-asset case. 
The developed algorithm leverages the exponential amplitude loading technique introduced in \cite{cibrario2024} to price multi-asset path-independent rainbow options. The authors introduced two methods to implement exponential amplitude loading: direct and integration.
We propose an improvement of the integration amplitude loading to enhance the postprocessing mapping of the expected value. This serves to reduce the overall accuracy requirements for the quantum algorithm compared to \cite{cibrario2024}.
Section \ref{subsec:intal} and \ref{subsec:improved_intal} discuss the integration amplitude loading technique independently of its application to option pricing. The symbols used herein are defined locally and do not correspond to those in other sections.
\subsection{Integration Amplitude Loading}
\label{subsec:intal} 
Given a quantum state $|x\rangle$ composed of $n$ qubits and interpreted as an integer, the integration amplitude loading technique introduced in \cite{cibrario2024} enables the encoding of a summation over a pre-loaded probability distribution into the amplitude of the state.
 As depicted in Figure \ref{fig:integrator}, it uses $n$ additional qubits prepared with the state $\sum_r\sqrt{g(r)}|r\rangle$ plus one auxiliary qubit and a comparator to create the following state:
\begin{equation}
\label{eqn:integrator}
    \sqrt{\sum_{r=0}^x|g(r)|}|x\rangle_n|\psi\rangle_n|1\rangle + \sqrt{\sum_{r=x+1}^{2^n-1}|g(r)|}|x\rangle_n|\psi_\perp\rangle_n|0\rangle
\end{equation}
where $|\psi\rangle_n$ and $|\psi_{\perp}\rangle_n$ are two orthogonal states. The comparator basically performs the integral of the function $g(r)$. 
To load an exponential function, $g(r)$ is defined as:
\begin{equation}
\label{eqn:g(r)}
    g(r) = \frac{e^{a r}}{Z}
\end{equation}
where $Z$ is the normalization factor ensuring the state in the $r$ register is properly normalized. In this specific context, $Z$ is equal to $\sum_i{e^{a i}}$.
 Loading this state requires only a set of 
 $n$ parallel $R_y$ rotations, each with angle: 
\begin{equation}
\label{eq:sp_exp_angles}
\theta_i = 2\arctan{\left(e^{\frac{a2^i}{2}}\right)},
\end{equation}
as shown in Figure \ref{fig:exponential_sp}.
With this exponential state preparation, the amplitude of the $|1\rangle$ state of the auxiliary qubit in \ref{eqn:integrator} becomes:

\begin{equation}
\label{eqn:exponential_integrator}
\sqrt{\frac{e^{a(x+1)}-1}{e^{a2^n}-1}}
\end{equation}

If the input $x$ is limited to only a subset of the possible states of its quantum register, the normalization factor $e^{a 2^n} - 1$ becomes excessive, as it is designed for the full range of $2^n$ states. Since only a restricted input domain is considered, this overestimation negatively affects the accuracy of subsequent estimates.

\subsection{Improving Integration Amplitude Loading}
\label{subsec:improved_intal}
We suggest a mitigation for the normalization issue of the method described in \ref{subsec:intal} by loading partial exponential states. By preparing the register $r$ with an exponential state, but only for the relevant range (assuming it is continuous), the normalization is optimal. Denoting the limits of integration by $x_0<x_1 \in [0, 2^n-1]$, loading the register $r$ with the state:
\begin{equation}
\label{eqn:partial_exponential_state}
\sum_{x_0}^{x_1}\sqrt{\frac{e^{ar}}{Z}}|r\rangle
\end{equation}
results in the following amplitude after integration:
\begin{equation}
\label{eqn:better_scaled_state}
\begin{cases}
0, & \text{if } x < x_0, \\[6pt]
\sqrt{\frac{e^{a(x+1)}-e^{ax_0}}{e^{a(x_1+1)}-e^{ax_0}}}, & \text{if } x_0 \le x \le x_1, \\[6pt]
1, & \text{if } x > x_1.
\end{cases}
\end{equation}.

\begin{figure}[ht]
    \centering
    \begin{subfigure}{1\linewidth}
        \centering
        \includegraphics[trim={40 0 0 0}, width=0.8\linewidth]{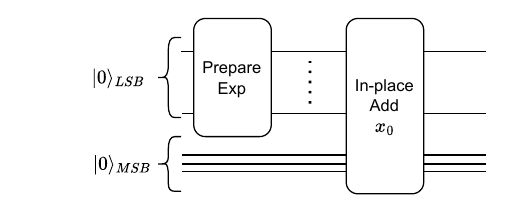}
        \caption{Implementation for interval size which is a power of $2$. The addition should be used in the case $x_0 \neq 0$.}
        \label{fig:exp_sp_2_power}
    \end{subfigure}
    \hfill
    \begin{subfigure}{1\linewidth}
        \centering
        \includegraphics[trim={10 0 0 0}, width=\linewidth]{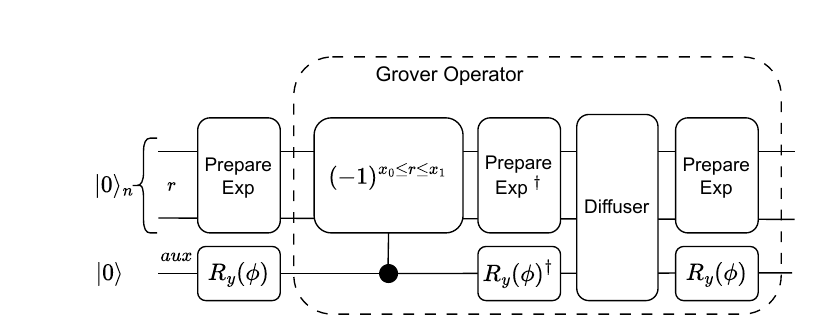}
        \caption{Implementation when interval size is not a power of $2$. The circuit uses the exact Amplitude Amplification scheme described in \cite{mcardle2022quantumstatepreparationcoherent}. The angle $\phi$ is chosen such that the amplification is exact. The auxiliary is freed by the amplification. If the interval $[x_0, x_1]$ does not contain most of the probability, it is possible to replace the block `Prepare Exp` with the module in \ref{fig:exp_sp_2_power} to use a single amplification round.}
        \label{fig:exp_sp_general}
    \end{subfigure}
    \caption{
    State preparation of exponential on a specific interval: $\sum_{x_0}^{x_1}\sqrt{\frac{e^{ar}}{Z}}|r\rangle$.}
    \label{fig:subfigures2}
\end{figure}

We now describe the procedure for preparing a partial exponential state, given by $\sum_{x_0}^{x_1}\sqrt{\frac{e^{ar}}{Z}}|r\rangle$ for a specified $x_0$ and $x_1$. Exponential state preparation for the entire domain of a quantum variable, as described in \cite{cibrario2024}, can be done using a single $R_y$ rotation on each qubit, as depicted in Figure \ref{fig:exponential_sp}. For a limited domain $[x_0, x_1]$, if the difference $x_1 - x_0$ is a power of $2$, it can be efficiently loaded by exponential state preparation on the LSB\footnote{Least Significant Bits (whereas MSB stands for Most Significant Bits).} of $|r\rangle$ followed by an in-place addition if needed, as shown in Figure \ref{fig:exp_sp_2_power}.
If the difference is not a power of $2$, then the partial exponential state preparation includes a complete exponential state preparation on all qubits, followed by a single round of exact amplitude amplification as described in \cite{Long_2001, mcardle2022quantumstatepreparationcoherent}. The Grover operator should use a suitable comparator as an oracle, as shown in Figure \ref{fig:exp_sp_general}.

In the context of option pricing, the exponential function is needed to shift from the log-return space to the return space, as the payoff is commonly expressed in terms of returns or asset prices. We highlight that the entire state preparation on the $r$ register can be parallelized with other parts of the quantum circuit, leaving only the comparator to be done serially.

\subsection{Autocallable Option Pricing}
\label{subsec:autocallable}
We now detail a complete description of our method for pricing a single-asset autocallable option. 
With the addition of some arithmetic modules, the approach is fully extensible to the case of the \textit{best-of} or \textit{worst-of} basket autocallable. 
We applied Algorithm \ref{alg:quantum_algorithm} to obtain a quantum circuit generating a state in the form of Equation \ref{eqn:payoff_ae} and used IQAE \cite{Grinko2021} to estimate the probability of the target qubit being in the $|1\rangle$ state. 
Based on the \textit{re-parameterization method} described in \cite{chakrabarti2021} we loaded standard Gaussians (one for each time step) and then used arithmetic to adjust the mean and variance. 
Log-returns for each time step could then be calculated by accumulation using an in-place adder. We applied all the necessary comparators in the log-return space to check the binary options and the put option conditions. Instead of transitioning to the return space using arithmetic, we used the integration amplitude loading described in Section \ref{subsec:improved_intal}. 

The estimation target for the whole algorithm is:
$$
    \bar{f} = E[
    f_1e^{-rt_1}i_1 + ... + f_je^{-rt_j}i_j + fe^{-rT}i_{j+1} + 0i_{j+2}]
$$
where $f_k$ is the $k$-th binary's payoff and $f$ is the put option non-zero payoff. The rescaling $e^{-rt}$ is necessary to discount the payoff, and it can be included directly in the loaded payoff.
$i_k$ are indicator functions defined for mutually exclusive events. The first is set to $1$ just in case the first binary is in the money. From the second to the $j$-th the indicators are set to $1$ only if the indexed binary is in the money and if all of the previous ones were out of the money. $i_{j+1}$ is set to $1$ if the put option is activated in the money, and $i_{j+2}$ is set to $1$ if the put option is activated out of the money. 

Because the amplitude estimation returns a number in the range $[0, 1]$, we needed to rescale the payoff values to this range. In order to take full advantage of the range, we mapped an amplitude of $0$ to the minimum payoff value, $(r_{T_{\min}} - K)V$, and an amplitude of $1$ to the maximum payoff, ${f_{\max}=\max_i\{f_ie^{-rt_i}\}}$. This mapping results in a normalization factor of $f_{\max} + (K - r_{T_{\min}})V$, where $r_{T_{\min}}$ is the minimum possible value of the return at time $T$. Note that in this mapping, a zero payoff requires loading a number greater than $0$ into the amplitude.

Loading the constant payoffs ($f_ie^{-rt_i}$ and the zero payoff) according to the mapping was achieved using a single $R_y$ on an indicator qubit. 

The non-zero payoff loading for the put option was done with the integration amplitude loading mentioned in \ref{subsec:improved_intal}. 
The reference variable was loaded with an exponential distribution in the range of the put option non-zero payoff, between the minimum log-return and the strike log-return $(\ln K)$. 
Finally, we performed an $R_y$ rotation on an additional indicator qubit to map the normalization factor from $(K - r_{T_{\min}})V$ to $f_{\max} + (K - r_{T_{\min}})V$. 
An alternative to re-mapping the put option non-zero payoff is to load the exponential reference state including the entire payoff range (considering binary payoffs). This approach avoids the use of the additional indicator qubit.

\begin{algorithm*}
  \caption{Quantum algorithm for  pricing autocallable options}
  \label{alg:quantum_algorithm}
  \begin{algorithmic}[1]
    \renewcommand{\algorithmicrequire}{\textbf{Input:}}
    \renewcommand{\algorithmicensure}{\textbf{Output:}}
    \Require Mean of historical log-return $\mu$. Standard deviation of historical log-return $\sigma$. Gaussian discretization step $ds$, number of discretization qubits $n$ and truncation value $s_{min}$. 
             Time step size $dt$.
             Total number of time steps $T$.
              Barrier $b$ and strike return $K$ for the put option. Amplitude-mapped payoff $\tilde{f}_i$ for binary $i$ with strike $k_i$. Amplitude-mapped zero payoff $\tilde{f}_z$.
    \State Load a Gaussian distribution for each time step:
            \begin{equation}
             |G\rangle_{Tn} = \bigotimes_{i=1}^{T}\mathit{G}|0\rangle_n = \bigotimes_{i=1}^{T}\left(\sum_{g=0}^{2^n-1} \sqrt{p(g_{i})}|g_{i}\rangle_n\right)  
            \end{equation}
            where $g_i$ is the $g$-th discretization point of the $i$ discrete standard Gaussian and $p(g_{i})$ is the respective probability.
    \State Prepare an accumulator register of size $m$ (sufficient to store cumulative log-returns up to time $T$) and compute the log-return for the first time step ($l_1$) using the operator $R_1$:

    \begin{equation}
             |G\rangle_{Tn}|0\rangle_m \xrightarrow{\mathit{R_1}} |G\rangle_{Tn} |l_1\rangle_m = |G\rangle_{Tn} |l(g_1)\rangle_m
            \end{equation}
    where the function $l(g)$ adjusts the mean and variance of the Gaussian sample:
    \begin{equation}
    \label{eqn:affine}
    l(g) = \mu \cdot dt + \sigma (g \cdot ds+s_{min}) \cdot \sqrt{dt}
    \end{equation}
            
\For {$t$ in $[1, \dots ,T]$}
        \If {$t\neq 1$}
            \State Compute the log-return $l_t$ at time $t$ through an operator $R_t$ by applying the function in Equation \ref{eqn:affine} to $g_t$, and then adding the log-return from the previous time step ($l_{t-1}$):
                \begin{equation}
                 |G\rangle_{Tn} |l_{t-1}\rangle_m
            \xrightarrow{\mathit{R_t}}
            |G\rangle_{Tn} |l_t\rangle_m =
            |G\rangle_{Tn} |l_{t-1} +l(g_t)\rangle_m
                \end{equation}
        \EndIf  
    \State Update the register tracking the barrier crossing event $|C_t\rangle_T = |c_1c_2...c_{T}\rangle$. The initial value for the register is $|C_0\rangle_T =|0\rangle_T$. The operator $\mathit{C_t}$ sets the $t$-th qubit to $|1\rangle$ if the barrier is crossed at time $t$ (i.e., $c_t = (l_{t} < \ln(b))$):
            \begin{equation}
            |G\rangle_{Tn} |l_t\rangle_m |C_{t-1}\rangle_T  \xrightarrow{\mathit{C_t}}|G\rangle_{Tn} |l_t\rangle_m |C_{t}\rangle_T
            \end{equation}
    \State    
    Update the register tracking the verification of binaries' conditions $|B_t\rangle_j = |b_0b_1...b_{j-1}\rangle$ initialized as $|B_0\rangle_j =|0\rangle_j$. Let $\{s_0,s_1...s_{j-1}\} \subset[1,T-1]$ be the times at which binaries expire.  $b_i$ is set to $1$ if the binary defined at time $t=s_i$ is in the money and if all of the previous ones were out of the money (i.e., $b_i = (l_{s_i} > \ln(k_i)) \land (b_h = 0 \quad \forall h<i )$). Let $\mathit{B_j}$ be the operator  that performs this update:
            \begin{equation}
            |G\rangle_{Tn} |l_t\rangle_m |C_{t}\rangle_T |B_{t-1}\rangle_j \xrightarrow{\mathit{B_j}}             |G\rangle_{Tn} |l_t\rangle_m |C_{t}\rangle_T|B_{t}\rangle_j
            \end{equation}
            \State Apply a controlled  $R_y$ rotation via $b_i$ using operator $\mathit{F_i}$ to update the amplitude of the target qubit  (with initial state $|\tilde{f}_0\rangle = |0\rangle$) with the binary payoff (if $b_i=1$):
            \begin{equation}
            |G\rangle_{Tn} |l_t\rangle_m |C_T\rangle_T |B_{t}\rangle_j|\tilde{f}_{t-1}\rangle \xrightarrow{\mathit{F_i}} |G\rangle_{Tn} |l_t\rangle_m |C_T\rangle_T |B_{t}\rangle_j|\tilde{f}_t\rangle = |G\rangle_{Tn}  |l_t\rangle_m |B_{t}\rangle_j (\sqrt{\tilde{f}_i} \, |1\rangle + \sqrt{1 - \tilde{f}_i} \, |0\rangle)
            \end{equation}
    \EndFor
    \State Apply a controlled operator $\mathit{F}$ under the following conditions: no binary option was in the money ($|B_{T}\rangle_j =|0\rangle_j$), and the barrier was crossed at least once ($|C_T\rangle_T \neq |0\rangle_T$), and the put is in the money ($l_T < ln(K)$). The operator performs an exponential amplitude loading (defined in Section \ref{subsec:improved_intal}) to encode the amplitude-mapped non-zero payoff for the put option ($\tilde{f}$) in the target qubit: \begin{equation}
            |G_T\rangle |l_t\rangle_m |C_T\rangle_T|B_{T}\rangle_j|\tilde{f}_T\rangle \xrightarrow{\mathit{F}} |G\rangle_{Tn}  |l_t\rangle_m|C_T\rangle_T|B_{T}\rangle_j (\sqrt{\tilde{f}(l_T)} \, |1\rangle + \sqrt{1 - \tilde{f}(l_T)} \, |0\rangle)
            \end{equation}
    \State Apply a controlled $R_y$ rotation using the operator $\mathit{F_z}$ to encode $\tilde{f}_z$ (the value needed to obtain a $0$ payoff after postprocessing) when none of the above conditions are met:
    \begin{equation}
            |G\rangle_{Tn} |l_t\rangle_m |C_T\rangle_T |B_{T}\rangle_j|\tilde{f}_T\rangle  \xrightarrow{\mathit{F_z}}|G\rangle_{Tn} |l_t\rangle_m|C_T\rangle_T |B_{T}\rangle_j (\sqrt{\tilde{f}_z} \, |1\rangle + \sqrt{1 - \tilde{f}_z} \, |0\rangle)
    \end{equation}
  \end{algorithmic} 
\end{algorithm*}

\section{Complexity Analysis}
\label{sec:complexity_analysis}
Because fault-tolerant implementations of the T gate are resource intensive, it is crucial to minimize the T-depth in quantum algorithm development. In this section, we analyze the T-depth complexity of the whole algorithm, while considering the required accuracy \cite{niemann2019t}. 
We reduce the total T-depth compared to other proposed implementations of the same application \cite{chakrabarti2021}, \cite{Stamatopoulos2024derivativepricing}.

All sources of error in the algorithm should be bounded by the desired probability error threshold, $\epsilon_{}$:
$$
\epsilon_{\text{truncation}}, \,\epsilon_{\text{discretization}}, \,\epsilon_{\text{approximation}}, \,\epsilon_{\text{arithmetic}}, \,\epsilon_{\text{amplitude\ loading}} \leq \epsilon_{}
$$
where $\epsilon_{\text{truncation}}$, $\epsilon_{\text{discretization}}$ and $\epsilon_{\text{approximation}}$ are the Gaussian preparation's truncation, discretization, and approximation errors, respectively, while  $\epsilon_{\text{arithmetic}}$ and $\epsilon_{\text{amplitude\ loading}}$ are the errors resulting from the arithmetic and amplitude loading modules respectively.
For a given error threshold, one must choose the appropriate errors that do not exceed the threshold.

We analyze the whole algorithm's complexity by decomposing it into consequential building blocks, summing each one's T-depth, and considering the required error and the rescaling.

The whole T-depth is as follows:

$$
 D_{\text{tot}} = \\ 
 (1 + 2N_{\text{IQAE}}) \cdot \left( 
 \max{(D_{\text{G}} + D_{\text{arith}}, D_{\exp})} \\ 
 + D_{\text{AL}} 
 \right) 
$$

where:
\begin{itemize}
    \item $N_{\text{IQAE}}$ is the total number of calls to the operator building the state in Equation \ref{eqn:payoff_ae}. It is multiplied by $2$ because it appears twice in the Grover operator. The value of $N_{\text{IQAE}}$ is chosen according to $\epsilon_{}$ by $N\sim 1/\epsilon_{}$.
    \item $D_{\text{G}}$ is the T-depth of a Gaussian preparation block. All Gaussian preparation blocks, one for each timestep and asset, are called in parallel. The overall error is therefore multiplied by $T$, the number of time steps and $d$, the number of assets. To ensure the error bounding of the parallel modules, $D_{\text{G}}$ requests an error of \\$\epsilon_{\text{approximation}}/(T\cdot d)$.
    \item $D_{\text{arith}}$ is the T-depth of the arithmetic module that accumulates the affine-transformed log-return as explained in the re-parameterization method and the different relevant comparators of the algorithm, as in \cite{chakrabarti2021}, as well as the constant payoffs calculation (for binary options and the zero payoff). 
    \item $D_{\exp}$ is the T-depth of the partial exponential state preparation for the integration method, which happens in parallel to the Gaussian preparation and arithmetic.
    \item $D_{\text{AL}}$ is the amplitude loading T-depth.
\end{itemize}

\subsection{Rescaling Analysis}
\label{subsec:compl_rescaling}
Rescaling is one of the critical parameters that affects the T-depth. A given payoff error, $\epsilon_{\text{payoff}}$, corresponds to the following error $\epsilon_{}$ in the probability domain: 
\begin{equation}
\label{eq:error_rescaling}
    \epsilon_{} = \frac{\epsilon_{\text{payoff}}}{R}
\end{equation}
where $R$ is the normalization factor applied in the post-processing stage, that amplifies $\epsilon_{}$ accordingly.

The rescaling factor for the integration method is:
\begin{equation}\label{eq:R}
    R=f_{\max} + (K - r_{T_{\min}})V
\end{equation}
where $f_{\max}$ is the maximal binary payoff; see Section \ref{subsec:autocallable}.

\renewcommand{\arraystretch}{1.5}
\begin{table*}[ht]
    \centering
    \begin{tabularx}{\textwidth}{|p{1.5cm}|p{3cm}|p{4.5cm}|X|} \hline 
         & Function&  T-depth& Description\\ \hline 
         $D_{R_y}$ & $R_y$ with error $\epsilon$&  $3 \log_2{\frac{1}{\epsilon}}$& According to \cite{ross2014optimal}\\ \hline 
         $D_{\CR_y}$ & $\CR_y$ with error $\epsilon$&  $6 \log_2{\frac{2}{\epsilon}}$& A $\CR_y$ can be implemented by $2$ consequent applications $R_y(\theta/2)$ and $2$ CX \\ \hline 
         $D_{R_z}$ & $R_z$ with error $\epsilon$&  $ \log_2{\frac{1}{\epsilon}}$& According to \cite{ross2014optimal}\\ \hline 
         $D_{\text{Toffoli}}$ & Toffoli&  3& According to \cite{draper2004logarithmic}\\ \hline 
         $D_{\text{MCX}}(n)$ & Muti Controlled X (MCX) with $n$ control qubits&  $14 \log_3 \frac{n}{2}+5$& The T-depth of the MCX is reduced using relative Toffoli gates \cite{maslov2016advantages}, and further optimized by the Classiq platform \cite{classiq_library}\\\hline
 $D_{\text{comparator}}(n)$ & Comparator of $n$ qubits& $(2 \log_2 (n)+ 9) D_{\text{Toffoli}}$&According to  \cite{draper2004logarithmic}\\\hline
 $D_{\text{C-comp}}(n)$ & Controlled Comparator of $n$ qubits& $D_{\text{comparator}}(n)+D_{\text{MCX}}(3)-D_{\text{Toffoli}}$ &A controlled comparator can be implemented by controlling just the middle Toffoli gate of the comparator (resulting in a CCCX). The T-depth can then be derived from the one of the comparator\\\hline
 $D_{\text{adder}}(n)$ & Adder of $n$ qubits & 
 $(2 \log_2 (n) + 5) D_{\text{Toffoli}}$
 & According to \cite{draper2004logarithmic}\\\hline
 $D_{\text{multiplier}}(n)$ &  Multiplier of $n$ qubits& 
 $n(D_{\text{adder}}(n)+6)$ & According to  \cite{quantum7010002}\\\hline
    \end{tabularx}
    \caption{T-Depth of Basic Building Blocks.}
    \label{tab:basic_depth}
\end{table*}

\subsection{Complexity and Error Analysis of Gaussian Preparation}
\label{subsec:compl_state_prep}
We now analyze the size of the Gaussian preparation blocks for each time step based on the payoff error. The sources of error in this block are the approximation error, truncation error, and discretization error.

The Gaussian preparation block must be chosen such that these sources of error do not exceed $\epsilon_{}$. This requires choosing a suitable distribution truncation and an appropriate number of qubits per block, $k$, to capture a multidimensional Gaussian distribution for $T$ time steps and $d$ assets. The distribution has $\sigma_{\max}$ as the maximal eigenvalue of the covariance matrix for a unit time.
Following \cite{chakrabarti2021} we define $w$, the number of $\sigma_{\max}$ we use for the truncation to ensure ${\epsilon_{\text{truncation}} \leq \epsilon_{}}$. We require that  ${P(r_T<r_{T_{\min}})\leq\epsilon_{\text{truncation}}/2}$; this condition determines the minimum return at time T as ${r_{T_{\min}}=\exp(\mu \Delta tT-w\sigma_{\max}\sqrt{\Delta t}T)}$, where $\Delta t$ is the unit time per sample point. Once we have $w$, we can use it to find $k$, ensuring $\epsilon_{\text{discretization}} \leq \epsilon_{}$.

\subsubsection{Finding the Required Truncation }
\label{subsubsec: truncation}
The challenge in finding $w$ relies on the fact that $\epsilon_{}$ depends on $R$ according to Equation \ref{eq:error_rescaling}. Additionally, $R$ depends on $r_{T_{\min}}$ (see Equation \ref{eq:R}), which, in turn, depends on $w$. 

The truncation error is the sum of all $2 d T$ tails of the truncated Gaussian distributions \cite{chakrabarti2021}:
\begin{equation}
    \epsilon_{\text{truncation}} = 2 d T e^{-w^2/2}
\end{equation}
One must find $w$ that satisfies:
\begin{equation}
    \epsilon_{\text{truncation}} \leq \epsilon 
=\frac{\epsilon_{\text{payoff}}}{f_{\max} + (K - r_{T_{\min}})V}
\end{equation}

\subsubsection{Gaussian Preparation Depth Analysis}
Once we find the truncation parameter $w$, for each Gaussian preparation block we can determine the number of qubits $k$ that discretizes the truncated domain while maintaining the discretization error $\epsilon_{\text{discretization}}$; following the approach described in \cite{chakrabarti2021}.

The Gaussian preparation circuit on $k$ qubits is a variational circuit with $L$ layers. Each layer contains an $R_y$ rotation on each qubit. Therefore, the T-depth of each $R_y$, $D_{R_y}$, depends on the desired approximation error by $\epsilon_{R_y}=\epsilon_{\text{approximation}}/(k\cdot T \cdot d)$.
Thus, the total T-depth will be \cite{chakrabarti2021}:
\begin{equation}
    D_{\text{G}} =  (L+1) \cdot D_{R_y}.
\end{equation}
\subsection{Arithmetic and Constant Payoff Modules Complexity}
\label{subsec:compl_arithmetic}
Following the Gaussian preparation, we compute the log-return evolution over time, verify payoff conditions, and compute the constant payoffs. The log-return evolution and condition evaluation are performed using arithmetic operations, while a combination of $\MCX$ and $\CR_y$ gates is used to store constant payoffs in the amplitude of the target qubit according to the relevant conditions. The T-depth of these operations ($D_{\text{arith}}$) can be constructed from the elementary building blocks in Table~\ref{tab:basic_depth}.

\subsection{Complexity of Amplitude Loading Module}
\label{subsec:compl_al}
Our integration amplitude loading module (see Section \ref{subsec:improved_intal}) is controlled to identify the condition for non-zero put payoff activation. The module consists of a partial exponential state preparation block followed by an integration comparator (see Figure~\ref{fig:integrator}). The entire module can be controlled by applying control only to the integration comparator, as the partial exponential state preparation operates solely on the $r$-register. Since the partial exponential state preparation can be executed in parallel with the rest of the circuit, the amplitude loading complexity is determined solely by that of the controlled comparator:

\begin{equation} D_{\text{AL}} = D_{\text{C-comp}}(m) \end{equation}

where $m$ is the number of qubits required to store the log-return at the final time. The complexity of the partial exponential state preparation, $D_{\exp}$, is derived as follows.

\subsubsection{Complexity of partial exponential state preparation }
The partial exponential state preparation module (depicted in Figure \ref{fig:exp_sp_general}) runs in parallel to the modules described in Sections \ref{subsec:compl_state_prep} and \ref{subsec:compl_arithmetic}.
Its depth depends on $m$ and on the $\epsilon_{}$ requirement:
\begin{equation}
    D_{\exp} =
     3D_{Ry} + D_{\text{MCX}}(m) + 2D_{\text{C-comp}}(m)
\end{equation}
The $3$ exponential state preparation blocks contain an $R_y$ rotation on each of the $m$  qubits as well as on the additional auxiliary qubit, totaling at $m+1$ qubits. Therefore, the T-depth, $D_{R_y}$, of each $R_y$ depends on the error by $\epsilon_{\text{amplitude loading}}/(m+1)$. The diffuser block contains an MCX, and there are two controlled comparators, one for each inequality; their depths are given in Table \ref{tab:basic_depth}.

\subsection{Comparison with State of the Art}
\label{subsubsec: complexity result}
We provide a T-depth analysis of our proposed autocallable pricing algorithm and compare it with an implementation based on QSP \cite{Stamatopoulos2024derivativepricing}. 
For a fair comparison, we consider the modules presented in Sections \ref{subsec:compl_state_prep} and \ref{subsec:compl_arithmetic} for both approaches, changing only the amplitude loading module, which constitutes the primary distinguishing feature between the two methods.
The problem parameters are chosen to be the same as in \cite{Stamatopoulos2024derivativepricing}, with $T=20$ timesteps, $d=3$ assets and $\epsilon=2\cdot10^{-3}$ total estimation error. 

For the integration amplitude loading method proposed in \cite{cibrario2024}, normalization diminishes any improvement obtained from amplitude loading.
Instead, the method presented in Section \ref{subsec:improved_intal} shows a significant improvement for the amplitude loading module, from $\sim2.1\cdot 10^3$ to $\sim40$ T-depth (considering the parallelization of the partial exponential state preparation module), without significantly altering the depth of the other modules.
Such a significant enhancement in the T-depth necessary for the payoff component suggests that further improvements to the overall T-depth are unlikely to come from optimizing the amplitude loading module alone.
However, it opens the door to further improvements in other components of the quantum algorithm.

\section{Experiments}
\label{sec:experiments}
This section presents the results of an extensive series of simulation experiments aimed at validating the proposed model. 
The input data for replicating such experiments can be found in Table \ref{tab:input_data}.
The implementation is intended to be a proof of concept for the integration methodology in the context of autocallable option pricing. 
The full algorithm for pricing autocallable options is based on the method described in \ref{subsec:intal} to avoid implementing different problem-dependent models required by the method reported in \ref{subsec:improved_intal}. 
However, a code base implementing the different strategies for loading the exponential state preparation is also available together with the whole algorithm \cite{repo} and can be used to adapt the implementation to obtain optimal rescaling, given the problem parameters.

We used the Classiq platform to build a high-level functional model and synthesize it to obtain the resulting gate-based quantum circuit \cite{vax2025qmod, Qmod_docs}.
The Classiq synthesis engine explores multiple implementations with the same functionality and returns an optimized circuit based on user-defined preferences, such as optimizing circuit depth, number of qubits, or two-qubit gate usage, or tailoring the circuit for specific hardware \cite{classiq-library2024}.

Currently available quantum computers have limited coherence times and their fidelity is not yet sufficient for large-scale circuits executions. This makes them unsuitable for validating the proposed algorithm. For this reason, we employed quantum simulators to validate our approach.
To enable circuit execution on simulators, we synthesized the circuit minimizing its width to reduce the number of qubits required.

\begin{table}[ht]
    \centering
    \caption{Model Parameters for Reproducibility}
    \label{tab:input_data}
    \begin{tabular}{l c}
        \hline
        \textbf{Parameter} & \textbf{Value} \\
        \hline
        Notional value ($V$) & 18\$ \\
        Time step ($dt$) & 1 year \\
        Number of time steps ($T$) & 3 \\
        Annual volatility ($\sigma$) & $0.2382$ \\
        Annual average log-return ($\mu$) & $0.1274$\\
        Put barrier ($b$) & 0.7 \\
        Put strike return ($K$)& 1 \\
        Risk-free rate ($r$) & $4\%$\\
        First binary option strike ($k_1$) & 1.1 \\
        Second binary option strike ($k_2$) & 1.1 \\
        First binary option payoff ($f_1$)& $2$ \\
        Second binary option payoff ($f_2$)& $5$ \\
        Standard Gaussian truncation ($s_{min}$) & $3$\\
    \end{tabular}
\end{table}

For our experiments, we considered an autocallable option with a single asset, three time steps, and two binary options expiring at the first two monitoring points.

The goal of the quantum algorithm is to improve the convergence rate compared to traditional Monte Carlo methods. Therefore, the ideal validation for the quantum circuit should consist in comparing its execution outcomes with those obtained using this classical methodology. However, a fair comparison is not feasible on current simulators because quantum circuits synthesized under the accuracy constraints outlined in Section \ref{sec:complexity_analysis} exceed the simulators’ capabilities. In fact, existing simulators can only handle quantum circuits with very limited width and depth.

To address this limitation, we built a set of classical models to experimentally demonstrate that it is possible to achieve convergence to the traditional Monte Carlo result by starting with a circuit executable on current simulators and then upgrading it, increasing the number of qubits allocated for Gaussian encoding and quantum arithmetic precision.

We conducted comparisons using (1) the traditional Monte Carlo methodology, (2) a Monte Carlo methodology relying on the same Gaussian discretization of the quantum algorithm, (3) a closed form calculation constrained by the same Gaussian discretization and (4) a closed form considering both the Gaussian discretization and the arithmetic precision of the quantum model. 

\begin{figure}
    \centering
    \includegraphics[width=\linewidth]{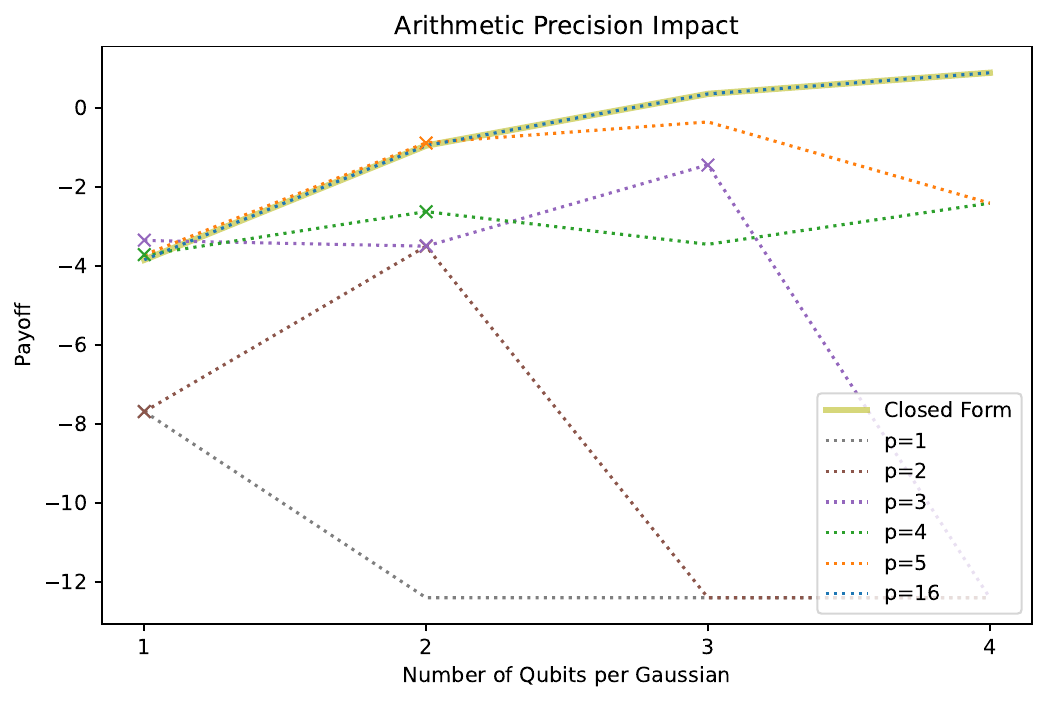}
    \caption{
    Expected payoff for the following classical models: (1) closed form with Gaussian discretization, (2) closed form with Gaussian discretization and explicit fixed point precision (where $p$ denotes the number of qubits used for the fractional part). The crosses 'x' represent results of the execution on simulators of the quantum circuit, synthesized with the given precision and discretization qubits. The configurations $[p=2, n_{\text{gaussian qubits}}=1]$, $[p=3, n_{\text{gaussian qubits}}=1]$ were executed on Classiq default simulator; $[p=4, n_{\text{gaussian qubits}}=1]$, $[p=3, n_{\text{gaussian qubits}}=2]$, $[p=4, n_{\text{gaussian qubits}}=2]$ were executed on the Nvidia simulator; while $[p=5, n_{\text{gaussian qubits}}=2]$ and $[p=3, n_{\text{gaussian qubits}}=3]$ were executed on LEONARDO's cluster.}
    \label{fig:run}
\end{figure}

The fourth benchmarking approach was essential to validate our quantum circuit results by comparing them with a classical model subject to the same constraints. In fact, we observed that the results of the classical algorithm fell within the confidence interval provided by the iterative version of the QAE \cite{Grinko2021}. We set the target confidence level choosing $\epsilon=0.001$ and $\alpha=0.002$.
 The crosses in Figure \ref{fig:run} represent the results of executions performed on exact shot-based simulators. For up to 25 qubits, we used the default Classiq simulator. For up to 30 qubits, we used the Nvidia simulator on a single GPU. For circuits exceeding 30 qubits, we executed the program on the LEONARDO HPC system  \cite{Turisini2024}, using the Cirq simulator on top of a cuQuantum appliance with multiple GPUs.

\begin{figure}
    \centering
    \includegraphics[width=\linewidth]{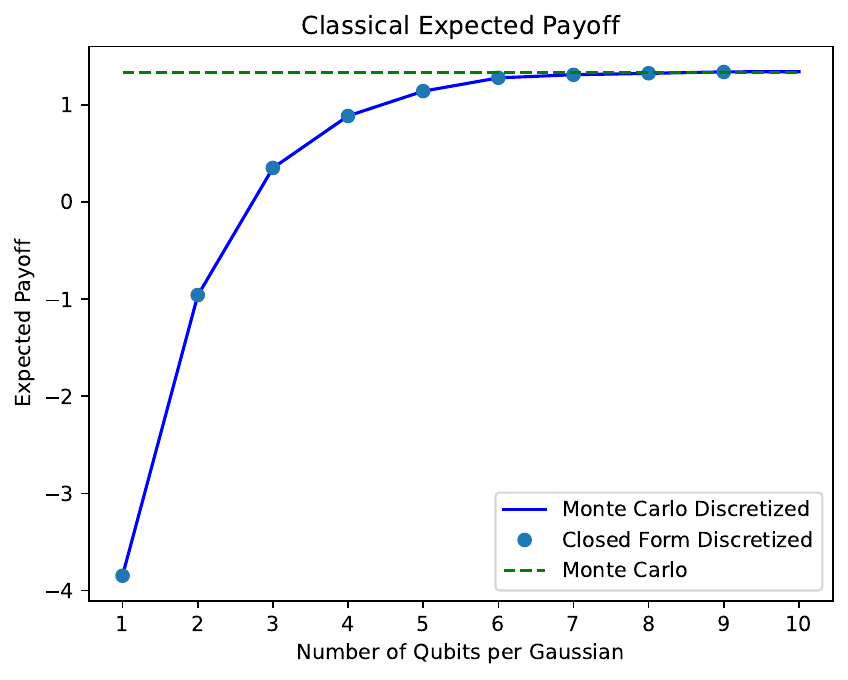}
    \caption{
    Expected payoff under the following methods: (1) traditional Monte Carlo simulation, (2) Monte Carlo simulation accounting for the number of discretization qubits for the Gaussian distribution, and (3) closed form calculation with the same discretization.}
    
    \label{fig:classical_mc_payoff}
\end{figure}

The third classical algorithm shows the behavior that the quantum circuit would exhibit if the number of precision qubits for digital encoding matched the classical machine precision. As shown in Figure \ref{fig:run}, increasing the number of precision qubits causes the fourth classical model (benchmarked with the quantum algorithm) to converge with the third. The third classical closed form algorithm, which still incorporates Gaussian discretization, becomes impractical as the number of discretization points and time instances increases, requiring prohibitive computational and memory resources. However, we can verify from Figure \ref{fig:classical_mc_payoff} that the results of a Monte Carlo algorithm considering the same Gaussian discretization overlap with the ones obtained with the closed form calculation. In the same figure, a verification of convergence of the two Monte Carlo methodologies (the discretized and non-discretized one) can also be observed. Through these tests, we had the opportunity to demonstrate the convergence of our results to those of the traditional Monte Carlo methodology, once it becomes possible to increase the number of qubits underlying the quantum circuits.

\section{Conclusion}
\label{sec:conclusions}
We presented a novel approach for pricing autocallable options. This includes a full end-to-end implementation using the Classiq platform and the Qmod language. 
Our methodology builds upon the integration-based exponential amplitude loading introduced in \cite{cibrario2024} and enhances it by not degrading the normalization factor.

Furthermore, we present a comprehensive complexity analysis to compare our approach with the state-of-the-art in the literature, particularly with the implementation based on Quantum Signal Processing in \cite{Stamatopoulos2024derivativepricing}. 
From this comparison, we saw a significant decrease in the necessary depth in terms of T-gates for the subroutine of amplitude loading.
Specifically, using our method we obtain a $\sim50$x reduction in a relevant setting.

We also presented the results of extensive simulation campaigns using, among others, the Nvidia simulator and the LEONARDO supercomputer.
These results validate our approach and illustrate how, by increasing the qubits allocated to arithmetic precision and Gaussian discretization, the behavior fits expected outcomes more effectively, enabling progressively closer values to those obtained with the classical Monte Carlo methodology.

The main challenge remains the considerable size of the circuits obtained even for small-scale examples, which necessitates the use of simulators, making potential tests on real hardware less interesting.

In the future, further optimization aimed at reducing the circuit size is desirable, as well as initiating a campaign of tests on real hardware that leverages error mitigation techniques to preserve the output. Additionally, the extension of the methodology to more complex volatility models could be explored.

\section*{Acknowledgment}
We acknowledge the ICSC for awarding this project access to the EuroHPC supercomputer LEONARDO, hosted by CINECA (Italy).
The authors thank Chani Sacharen for the valuable help in editing and proofreading the paper, ensuring clarity and coherence throughout the manuscript.

\bibliographystyle{IEEEtran}
\bibliography{biblio}
\end{document}